\documentclass[
 preprint,
 longbibliography,
 amsmath,
 amssymb,
 aps,
 nofootinbib]{revtex4-1}

\usepackage{graphicx}
\usepackage{dcolumn}
\usepackage{bm}
\usepackage{hyperref}
\usepackage{gensymb}
\usepackage{siunitx}
\usepackage{xcolor}
\usepackage{footmisc}

\begin{document}

\preprint{APS/123-QED}

\title{Dynamics of Tipping Cascades on Complex Networks}

\author{Jonathan Kr\"onke$^{1,2}$}
\author{Nico Wunderling$^{1,2,3,}$\footnote{\label{note1} Correspondences should be addressed to: donges@pik-potsdam.de or wunderling@pik-potsdam.de}}
\author{Ricarda Winkelmann$^{1,2}$}
\author{Arie Staal$^{4}$}
\author{Benedikt Stumpf$^{1,5}$}
\author{Obbe A. Tuinenburg$^{4,6}$}
\author{Jonathan F. Donges$^{1,4,}$\footref{note1}}
\affiliation{%
 $^1$Earth System Analysis, Potsdam Institute for Climate Impact Research, Member of the Leibniz Association, 14473 Potsdam, Germany, EU
}%
\affiliation{%
 $^2$Institute of Physics and Astronomy, University of Potsdam, 14476 Potsdam, Germany, EU
}
\affiliation{%
 $^3$Department of Physics, Humboldt University of Berlin, 12489 Berlin, Germany, EU
}%
\affiliation{%
 $^4$Stockholm Resilience Centre, Stockholm University, 10691 Stockholm, Sweden, EU
}%
\affiliation{%
 $^5$Department of Physics, Free University Berlin, 14195 Berlin, Germany, EU
}%
\affiliation{%
 $^6$Copernicus Institute, Faculty of Geosciences, Utrecht University, Utrecht, Netherlands, EU
}%

\begin{abstract}
Tipping points occur in diverse systems in various disciplines such as ecology, climate science, economy or engineering. Tipping points are critical thresholds in system parameters or state variables at which a tiny perturbation can lead to a qualitative change of the system. Many systems with tipping points can be modeled as networks of coupled multistable subsystems, e.g. coupled patches of vegetation, connected lakes, interacting climate tipping elements or multiscale infrastructure systems. In such networks, tipping events in one subsystem are able to induce tipping cascades via domino effects. Here, we investigate the effects of network topology on the occurrence of such cascades. Numerical cascade simulations with a conceptual dynamical model for tipping points are conducted on Erd\H{o}s-R\'enyi, Watts-Strogatz and Barab\'asi-Albert networks. {Additionally, we generate more realistic networks using data from moisture-recycling simulations of the Amazon rainforest and compare the results to those obtained for the model networks. We furthermore use a directed configuration model and a stochastic block model which preserve certain topological properties of the Amazon network to understand which of these properties are responsible for its increased vulnerability.} We find that clustering and spatial organization increase the vulnerability of networks and can lead to tipping of the whole network. These results could be useful to evaluate which systems are vulnerable or robust due to their network topology and might help to design {or manage} systems accordingly.
\end{abstract}

\pacs{Valid PACS appear here}
\maketitle


\section{\label{sec:intro}Introduction}
In the last decades the study of tipping elements has become a major topic of interest in climate science. Tipping elements are subsystems of the Earth system that may pass a critical threshold (tipping point) at which a tiny perturbation can qualitatively alter the state or
development of the subsystem \cite{Lenton2008}. However, tipping points also occur in various complex systems such as systemic market crashes in financial markets \cite{May2008}, technological innovations \cite{Herbig1991} or shallow lakes \cite{Scheffer2007} and other ecosystems \cite{Scheffer2001}. Understanding their dynamics is thus crucial not only for climate science but also for other disciplines that use complex systems approaches.\par
Many tipping elements are not independent from each other \cite{Brummit2015}. In such cases, if one tipping element passes its tipping point, the probability of tipping of a second tipping element is often increased \cite{Kriegler2009}, yielding the potential of tipping cascades \cite{Steffen2018} via domino effects with significant potential impacts on human societies in the case of climate tipping elements \cite{Cai2016}. In this study, we investigate the dynamics of complex networks of interacting tipping elements. A tipping element is described by a differential equation based on the normal form of the cusp catastrophe which exhibits fold-bifurcations and hysteresis properties. The interactions are accounted for by linear coupling terms. Many environmental tipping points can be described as fold bifurcations \cite{Lenton2013} and prototypical conceptual models that exhibit fold bifurcations have been developed for the Thermohaline Circulation \cite{WrightStocker1991}, the Greenland Ice Sheet \cite{LevermannWinkelmann2016}, or tropical rainforests \cite{Staal2018b} among others. Coupled cusp catastrophes have already been studied in detail for two or three subsystems \cite{Abraham1991,Brummit2015, Klose2019} or in combination with Hopf bifurcations \cite{Dekker2018}. On the other hand, threshold models for global cascades on large random networks have been investigated \cite{Watts2002}. \par 
{Here, we study cascades in complex systems with continuous state space that are moderate in size, yet large enough for statistical properties of the complex interaction networks to become relevant. Cascades in complex systems with continuous state space have been investigated for example for power grids  \cite{Yang2017,Schaefer2018}. We use a paradigmatic coupled hysteresis model based on the normal form of the cusp catastrophe. {Employing different network topologies such as Erd\H{o}s-R\'enyi-, Watts-Strogatz- and Barab\'asi-Albert-networks as well as networks generated from moisture-flow data of the Amazon rainforest, we investigate the effect of topological properties of the network.} We find that networks with a large average clustering coefficient are more vulnerable to cascading tipping and discuss how this is connected to the occurence of small-scale motifs such as direct feedback and feed-forward loops. We consistently observe networks with spatial organization like small-world or the Amazon networks are more vulnerable than strongly disordered networks.}
\section{\label{sec:method}The Model}
\subsection{System}
\begin{figure}
    \centering
    \includegraphics[width=0.6\textwidth]{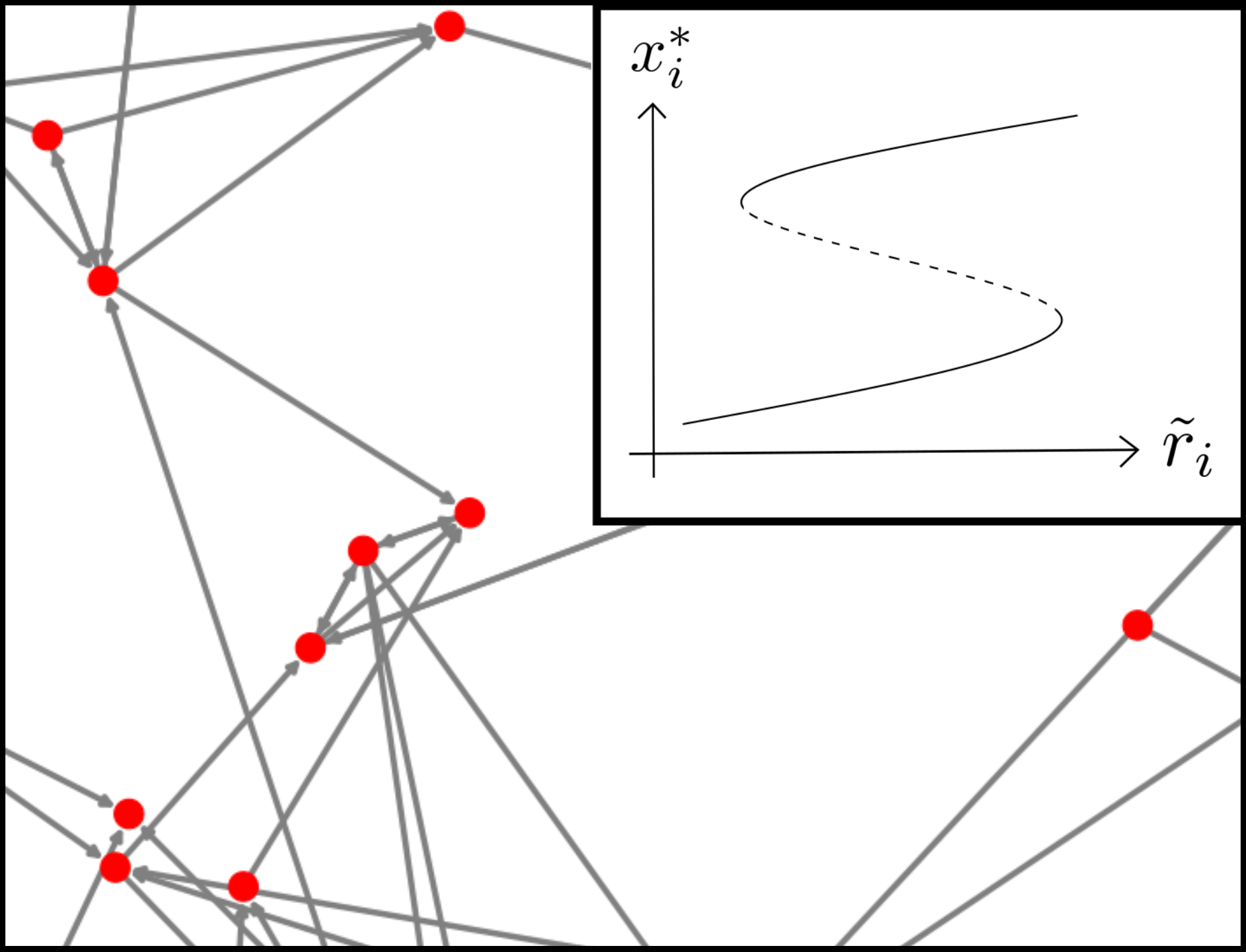}
    \caption{Illustration of a tipping network. Each node represents a tipping element with a corresponding state variable $x_i$. A directed link corresponds to a positive linear coupling with strength $d$. The effective control parameter $\tilde{r}_i$ of a node depends on the state of the nodes it is coupled to. The equilibria with respect to the effective control parameter are qualitatively illustrated in the box.}
    \label{fig:graph}
\end{figure}

In our conceptual model, a tipping element is represented by a (real) time-dependent quantity $x(t)$ that evolves according to the autonomous ordinary differential equation
\begin{equation}
    \frac{dx}{dt}=-a(x-x_0)^3+b(x-x_0)+r,
    \label{eq:dgl}
\end{equation}
where $r$ is the control parameter and $a,b>0$. The parameters $a$ and $b$ control the strength of these effects respectively and $x_0$ controls the positon of the system on the x-axis. The equation has thus one stable equilibrium for $|r|>r_\mathrm{crit}$ and a bistable region for $-r_\mathrm{crit}<r<r_\mathrm{crit}$ (see the bifurcation diagram depicted in the box in Fig.~\ref{fig:graph}). \par
We describe the characteristic behaviour of Eq.~\ref{eq:dgl}: If the system state is initially in the lower stable equilibrium ($x\approx0$) and $r$ is slowly increased, eventually at $r=r_{crit}$ a tipping point is reached and a critical transition to the upper stable equilibrium ($x\approx1$) occurs. If $r$ is afterwards decreased, the system state stays on the upper branch and only at $r=-r_{crit}$ tips down to the lower branch again. Equation~\ref{eq:dgl} is a minimal model for ecosystems with alternative stable states and hysteresis \cite{Scheffer2001} but can as well be used to conceptualize other systems with similar properties such as the Thermohaline Circulation and ice sheets \cite{Stommel1961, LevermannWinkelmann2016}.
\par
Next, we consider a directed network of $N$ interacting tipping elements as a linearly coupled system of ordinary differential equations
\begin{align}
    \frac{dx_i}{dt}=-a(x_i-x_0)^3+b(x_i-x_0)+\underbrace{r_i + d\sum_{j=1,j\neq i}^N a_{ij}x_j}_{\tilde {r}_i(x_1,x_2,...,x_N)},
    \label{eq:system}
\end{align}
where $d>0$ is the coupling strength and
\begin{equation}
    a_{ij}=
    \begin{cases}
    1, & \text{if there is a directed link from element $i$ to $j$}\\
    0, & \text{otherwise}
    \end{cases}.
\end{equation}
For simplicity, we use the same parameters $a$ and $b$ for all tipping elements in the network. An illustration of such a system with several tipping elements is depicted in Fig.~\ref{fig:graph}. Similar systems have already been studied with diffusive coupling focusing on hysteresis effects \cite{Eom2018}. \par
We briefly review the behaviour of two tipping elements with unidirectional coupling ($X_1 \rightarrow X_2$) \cite{Brummit2015}. The elements of the adjacency matrix are $a_{21}=1$ and $a_{12}=0$ which means that element 1 has an effect on element 2 but there is no effect in the other direction. As $r_1$ is slowly increased, it approaches its tipping point at $r_\mathrm{crit}$ and eventually tips from $x_{-}$ to $x_{+}$. The effective control parameter $\tilde{r}_2$ is thus increased by $\Delta \tilde{r}= d(x_{+}-x_{-})$. For $r_2=0$, a tipping event in the second element is induced if $\Delta \tilde{r}>r_\mathrm{crit}$ and therefore if the coupling strength exceeds a critical threshold of $d_\mathrm{c}=\frac{r_\mathrm{crit}}{x_{+}-x_{-}}$.
\subsection{Network Models}
To investigate the effect of the network topology on tipping cascades {we use different network models: We use three well-known models, the Erd\H{o}s-R\'enyi model (ER) \cite{Bollobas2001}, the Watts-Strogatz model (WS) \cite{Watts1998} and the Barab\'asi-Albert model (BA) \cite{Barabasi1999}. We slightly extend the two last models such that we are able to generate and compare directed networks with controllable average degree ${\langle k\rangle=\langle k_{in} + k_{out}\rangle}$. Furthermore, we use models to control the reciprocity and average clustering coefficient as well as a directed configuration model and a stochastic block model. All network models are shortly discussed in the following paragraphs:}\par
(i) The ER model is a simple random network model, where a directed link between two elements $i$ and $j$ is added with probability $p$. {The resulting average degree is $\langle k \rangle \approx p(N-1)$}. \par
(ii) The WS model is usually used to generate networks with large clustering coefficients, but small average path lengths to resemble the small-world phenomenon \cite{Milgram1967}. {We implement a directed WS model as follows: Initially, a regular network is generated where each node $i$ is connected in both directions to its $m$ nearest neighbors, e.g., nodes $i+1,i-1,...,i+\frac{m}{2},i-\frac{m}{2}$. Therefore, $m$ has to be an even integer and the average degree of the resulting regular network is equal to $m$. In order to generate networks with arbitrary average degree, $m$ is chosen such that the average degree of the resulting regular network is larger than the desired average degree. Then, until the average degree of the network matches the desired average degree, links are randomly deleted. Finally, each of the remaining links is rewired with probability $\beta$, similar to the usual WS model \cite{Watts1998}.} With increasing rewiring probability $\beta$ the generated network becomes more and more random and its properties approach the properties of ER networks for $\beta \rightarrow 1$. \par
(iii) The BA model is used to generate scale-free networks, i.e. networks with a power-law degree distribution. {We implement a directed BA model as follows: We start with two bidirectionally coupled nodes. Every additional node is in both directions connected to an already existing node $i$ with probability $p=\frac{k^\mathrm{in}_i+k^\mathrm{out}_i}{\sum_{m,n} a_{mn}}$. When the specified network size $N$ is reached, the average degree $\langle k\rangle \approx \frac{\sum_{m,n} a_{mn}}{N}$ is compared to the desired average degree. If the average degree is smaller than the desired average degree, links between randomly selected nodes $i$ and $j$ are added with probability $p=\frac{k^\mathrm{in}_i+k^\mathrm{out}_i+k^\mathrm{in}_j+k^\mathrm{out}_j}{2\sum_{m,n} a_{mn}}$ until the average degree matches the desired average degree. Else, if the average degree is greater than the desired average degree, links are randomly deleted as in the WS model.}\par
{(iv) To generate networks with arbitrary reciprocity $R$, we initially generate an ER network where all links are reciprocal ($R=1$). Afterwards, links are randomly chosen and rewired until the desired reciprocity is achieved.}\par
{(v) The procedure to generate networks with arbitrary average clustering coefficent $\mathcal{C}$ is similar. Initially a network with only reciprocal triangles between three randomly chosen nodes is generated. Afterwards links are randomly chosen and rewired again until the desired average clustering coefficient is achieved. That way, we are able to generate networks with an average clustering coefficient between $\mathcal{C}=0.05$ and $\mathcal{C}=0.35$. Note that the reciprocity is also large for networks with a large average clustering coefficient.}\par
{(vi) A directed configuration model can be used to generate networks with arbitrary average in- and out-degree. Links are randomly assigned to node pairs where the corresponding in- and out-degree has not been reached before \cite{Newman2001}.}\par
{(vii) Finally, stochastic block models (SBM) are used to generate networks with community structures. For each (directed) combination of communities there is a seperate link probability which is usually high within the community and small between two different communities \cite{Holland1983}.}
\subsection{\label{sec:method}Simulation Procedure}
\begin{figure}
    \centering
    \includegraphics[width=0.6\textwidth]{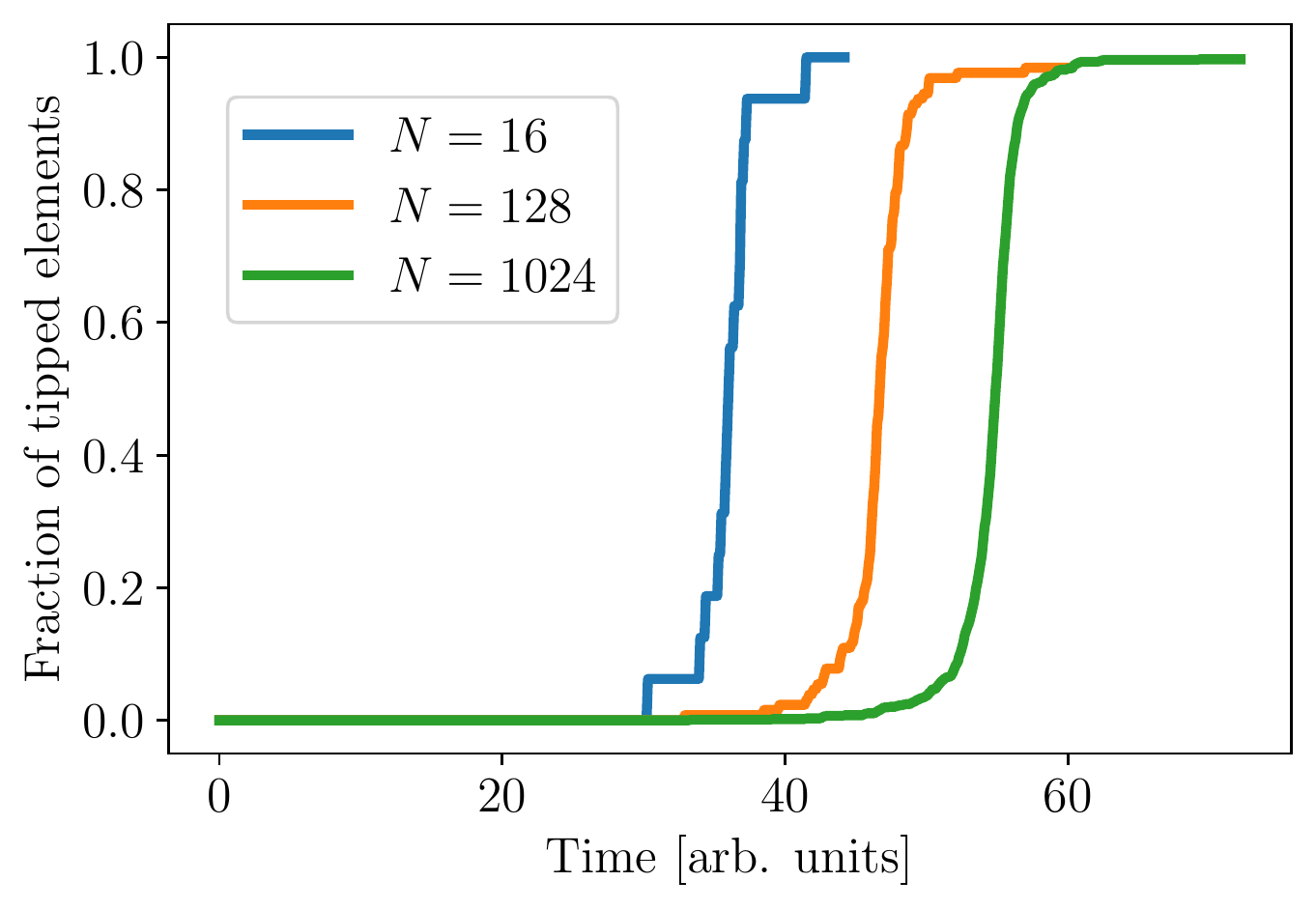}
    \caption{{Cascade simulations on ER-networks with different sizes, an average degree of $\langle k \rangle \approx 5$ and a coupling strength of $d=0.2$. The time evolution of the fraction of tipped elements is shown.}}
    \label{fig:tipped}
\end{figure}

We use the system given in Eq.~\ref{eq:system} and conduct cascade simulations on different network topologies. The parameters of the equation are chosen such that $r_\mathrm{crit}=0.183$ and for $r=0$ the two stable equilibria are $x_{-}=0$ and $x_{+}=1$ for all elements. The resulting parameters are $a=4$, $b=1$ and $x_0=0.5$.  Consider a network with $N$ tipping elements and a topology, that is described by the adjacency matrix $A=(a_{ij})$. Initially, $r_i=0$ and $x_i=0$ for all $i=1,...,N$. The algorithm of a cascade simulation is the following:
\begin{enumerate}
    \item Choose a random starting node $m$ of the network.
    \item {Slowly increase $r_m$ ($r_m \rightarrow r_m + \Delta r$).}
    \item {Let the system equilibrate, e.g., integrate the ODE system until $\dot{x}_i<\varepsilon$ for all $i=1,...,N$.}
    \item Check if at least one element tipped. If not jump back to step 2. Otherwise, count the number of tipped elements.
\end{enumerate}
We normalize the number of tipped elements to the number of nodes that can be reached on a directed path from the starting node (the size of the out-component), where we do not take the starting node into account. We call the resulting number cascade size $L$. The ODE system was {integrated} with the function \texttt{scipy.integrate.odeint} from the \texttt{scipy} python package \cite{Oliphant2007}. In all simulations, $\Delta r=0.01$  and $\varepsilon= 0.005$ was used. Examples of tipping cascades with size $L=1$ are shown in Fig.~\ref{fig:tipped} for ER networks with different size $N$.
\section{\label{sec:results}Results and Discussion}
\subsection{Cascades on Generic Network Topologies}
\begin{figure}
    \centering
    \includegraphics[width=0.6\textwidth]{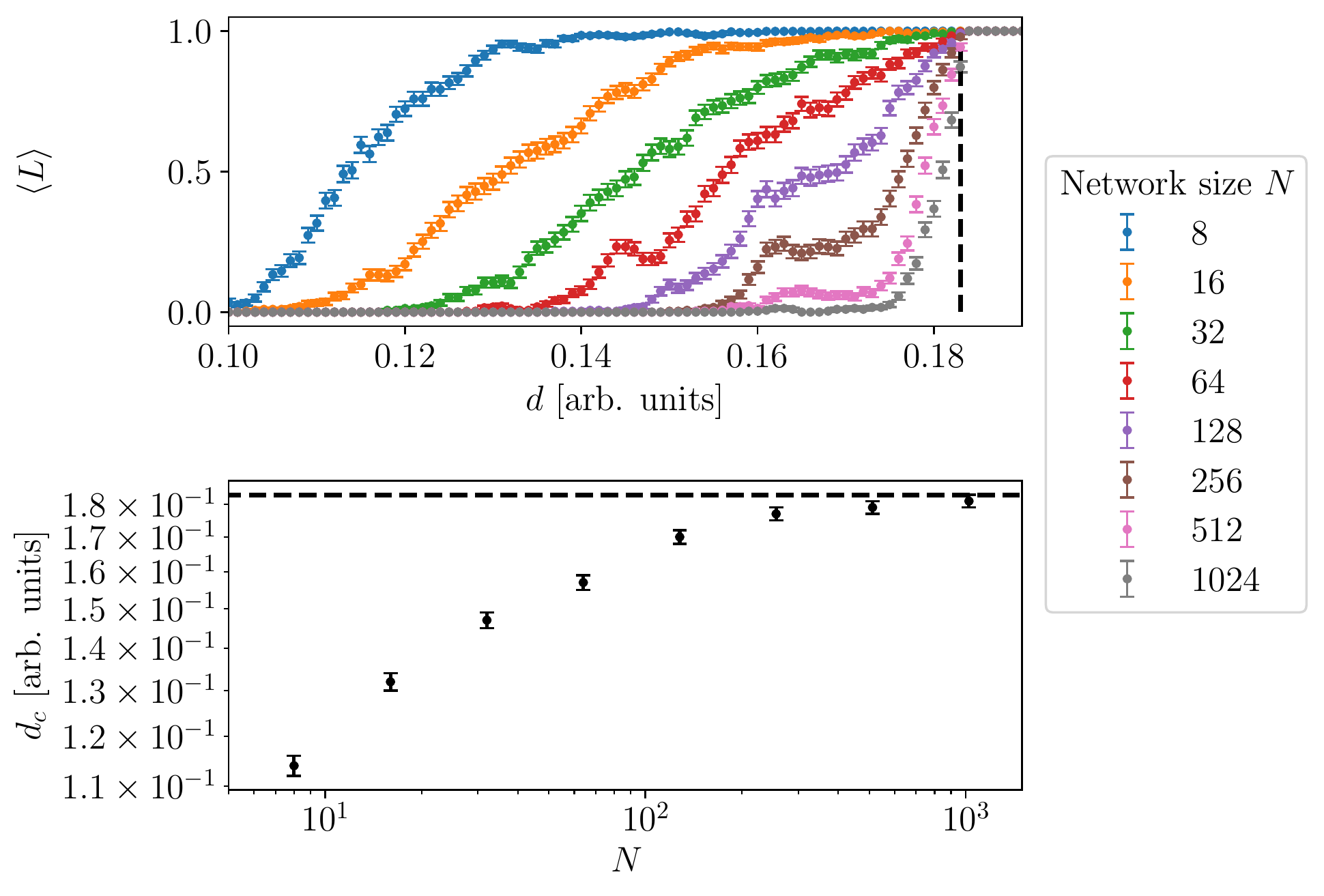}
    \caption{{Network size dependency of critical coupling strength in ER-networks with $\langle k\rangle \approx 5$. In the upper panel, the average cascade size with respect to the coupling strength in the transition region is shown. Each average is calculated from $100$ cascade simulations on different randomly generated networks with $N=100$. The error bars indicate the standard error. In the lower panel, the approximate critical coupling strength (coupling strength where $\langle L \rangle \approx 0.5$) with respect to network size $N$ is shown. The dashed line indicates the critical coupling strength $d_c \approx r_{crit}=0.183$ for a simple unidirectional coupling of two elements.} }
    \label{fig:size_dep}
\end{figure}

We start with cascade simulations on networks generated with the ER model. For any parameter combination we generate $100$ different networks and simulate one cascade on each network. We use the average cascade size from these simulations as a measure of the vulnerability of the corresponding network structure ranging from robust ($\langle L \rangle = 0$) to highly vulnerable ($\langle L \rangle = 1$) networks. The dependence of the average cascade size with respect to the coupling strength is shown in the upper panel of Fig.~\ref{fig:size_dep} for random networks with a fixed average degree $\langle k\rangle \approx 10$. For low coupling strengths ($d\lesssim 0.1$) the network is not affected by the externally induced tipping of one element and the average cascade size remains zero. With increasing coupling strength, a transition robust to vulnerable networks is observed. From the analysis of the unidirectional system, a sharp transition at $d\approx r_\mathrm{crit}$ would be expected for all networks. However, only for $N \rightarrow \infty$ the transition becomes more and more steep and approaches $r_\mathrm{crit}$. For networks of finite size, the onset of the transition is shifted to lower coupling strengths with decreasing network size. {We hypothesize that the reason for this is two different effects:} The first effect is the destabilization of the system by feedback loops ($X_1 \leftrightarrows X_2$) which can lead to a decrease of the tipping point $r_\mathrm{crit}$ of certain nodes. The second effect is due to the gradual change of the state of a tipping element $X_3$ that is coupled to another element ($X_1 \rightarrow X_3$). When the element $X_1$ tips, the state of the element $X_3$ will be slightly altered even if it does not tip. If it is coupled to another element $X_2$ however ($X_2 \rightarrow X_3$), the effective control parameter of element $X_3$ will be slightly increased by an increment of the order $\Delta \tilde{r} \sim d^2$. Therefore an additional indirect coupling with one intermediate node, called feed-forward loop, will decrease the critical coupling strength $d_\mathrm{c}$ of the target node. With this we can explain the size dependency of the transition which is shown in the lower panel of Fig.~\ref{fig:size_dep}. With increasing network size while fixing the average degree, the relative density of these motifs decreases and for $N\rightarrow \infty$, the destabilizing effect of the motifs vanishes.\par
\begin{figure}
    \centering
    \includegraphics[width=0.6\textwidth]{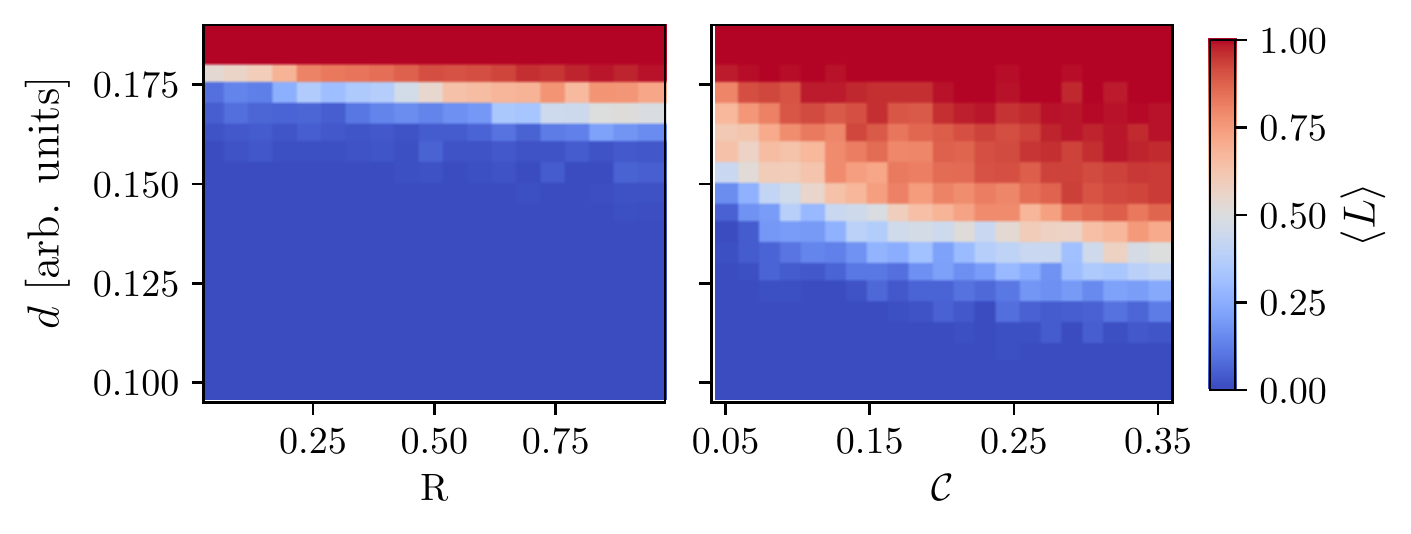}
    \caption{{Dependence of the transition region on the reciprocity $R$  (left panel) and on the clustering coefficient $\mathcal{C}$ (right panel). Each average is calculated from $100$ cascade simulations on different randomly generated networks with $N=100$.}}
    \label{fig:phases_rec_cl}
\end{figure}

{To test this hypothesis, cascade simulations on networks with different reciprocities and average clustering coefficients are conducted. Simulation results for different reciprocities $R$ can be seen in the left panel of Fig.~\ref{fig:phases_rec_cl}. As expected, for networks with high reciprocity, the transition region is shifted to lower coupling strengths. As can be seen, however, the dependence on the reciprocity is rather weak. Simulation results for networks with different average clustering coefficient $\mathcal{C}$ are shown in the right panel of Fig.~\ref{fig:phases_rec_cl}. It can be clearly seen that the vulnerability to tipping cascades is significantly increased for high average clustering coefficients. There are eight motifs that contribute to the average clustering coefficient in a directed network, two (indirect) feedback loops and six feed-forward loops \cite{Milo2002}. We suspect that the effect of indirect feedback loops is smaller than the effect of direct feedback loops for $d<1$. Therefore, we conclude that feed-forward loops are mainly responsible for the increased vulnerability of networks with large average clustering (see Fig.~\ref{fig:phases_rec_cl}).}\par
\begin{figure}
    \centering
    \includegraphics[width=0.6\textwidth]{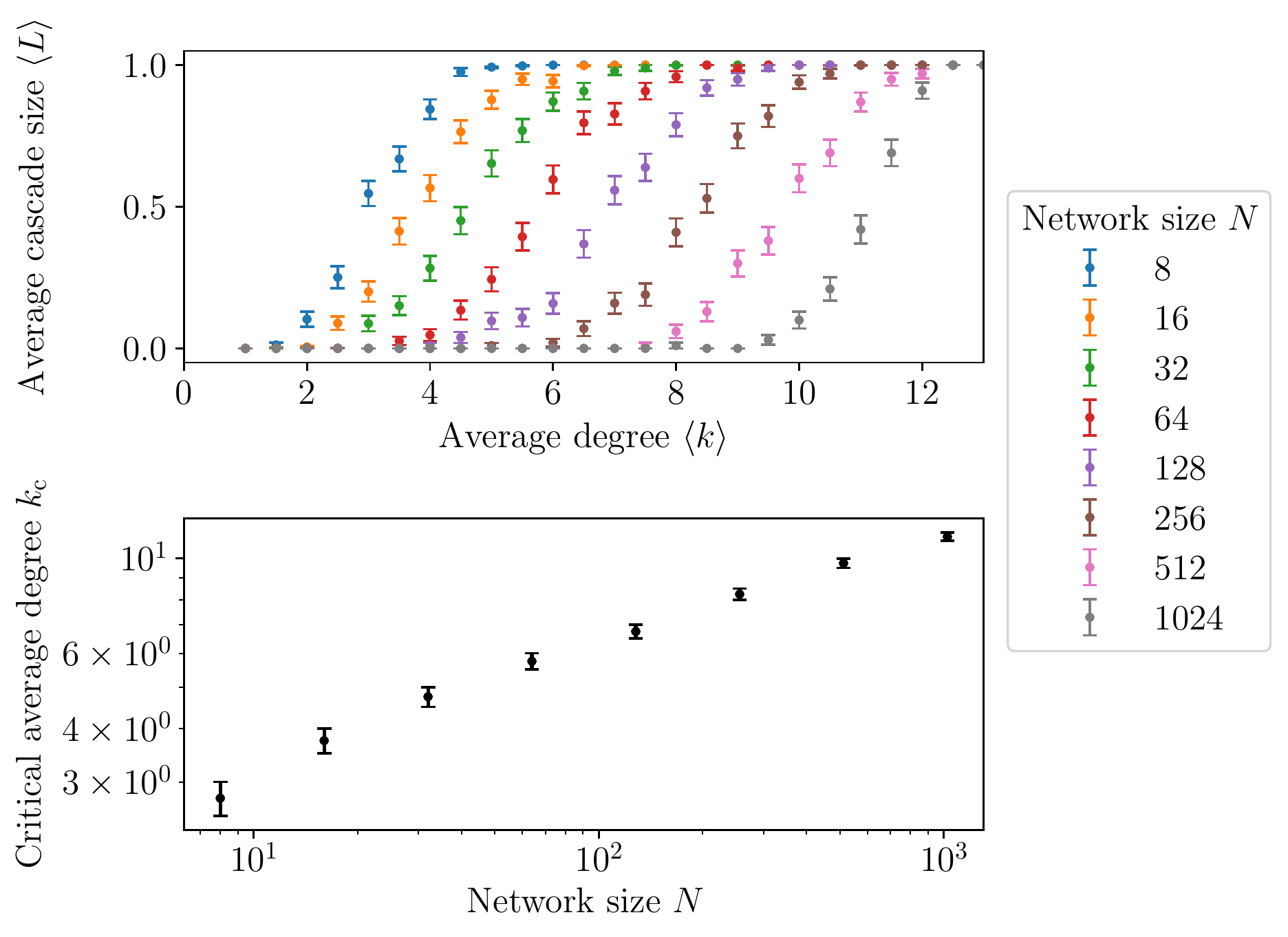}
    \caption{Network size dependency of critical average degree $k_\mathrm{c}$ in ER-networks with $d=0.15$. In the upper panel, the average cascade size with respect to the average degree in the transition region is shown. {Each average is calculated from $100$ cascade simulations on different randomly generated networks with $N=100$.} In the lower panel, the approximate critical average degree (average degree where $\langle L \rangle \approx 0.5$) with respect to network size $N$ is shown.}
    \label{fig:size_dep_2}
\end{figure}

We also observe a transition of the average cascade size when the coupling strength is held constant at $d=0.15$ and the average degree is varied (Fig.~\ref{fig:size_dep_2}). In that case the transition is shifted to higher average degrees when the network size increases, because a higher average degree is necessary to yield the same relative density of destabilizing motifs. \par
\begin{figure}
    \centering
    \includegraphics[width=0.6\textwidth]{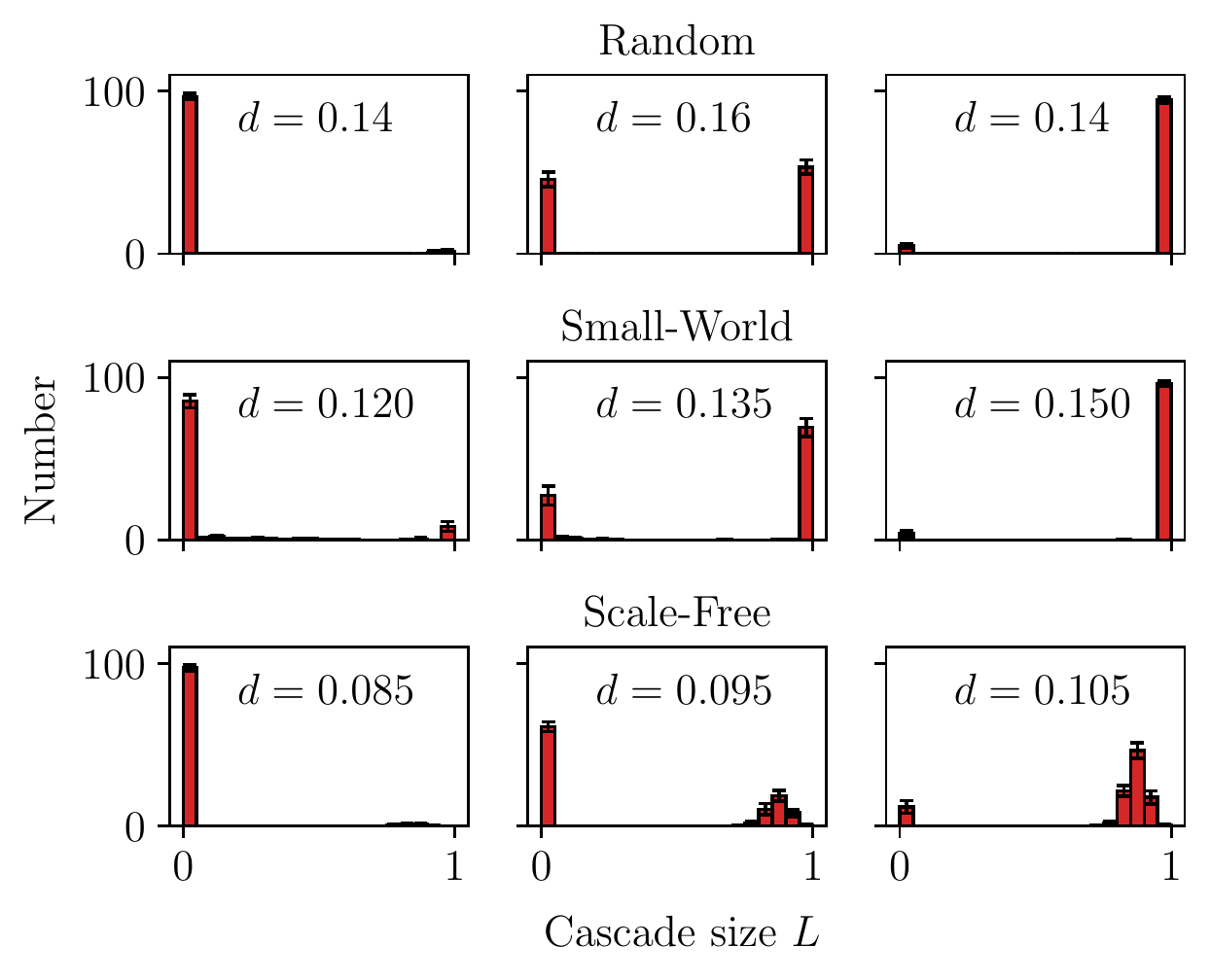}
    \caption{{Distributions of cascade sizes $L$ for different network topologies. A random topology generated with the ER model (first row), a small-world topology generated with the WS model (second row) and $\beta=0.1$ and a scale-free topology generated with the BA model (third row). Each distribution is an average of ten distributions with $100$ cascade simulations on different networks with $N=100$ and $\langle k \rangle \approx 5$. The error bars indicate the standard deviation across the ten distributions. Three different coupling strengths for each network topology are shown: one where almost no cascades occur; one where in about half of the simulations cascades are triggered; and one where in almost all simulations cascades are triggered.}}
    \label{fig:dist}
\end{figure}

Cascade distributions for $\langle k\rangle\approx 5$ and selected coupling strengths at the onset, in the center and at the end of the respective transition region are shown in Fig.~\ref{fig:dist}. We find a bimodal distribution of very small cascades ($L\approx 0$) and very large cascades ($L\approx 1$). {For networks with small-world and scale-free topology generated with the WS model with $\beta=0.1$ and the BA model, respectively, we observe similar transitions of the average cascade size. For the scale-free topology, the large cascades are distributed around an average size $\langle L \rangle < 1$.} This can be explained by the preferential attachment mechanism. Through this mechanism a large number of weakly connected elements develop which can only be tipped when the coupling strength is very large ($d\gtrsim r_\mathrm{crit}$).\par
\begin{figure}
    \centering
    \includegraphics[width=0.7\textwidth]{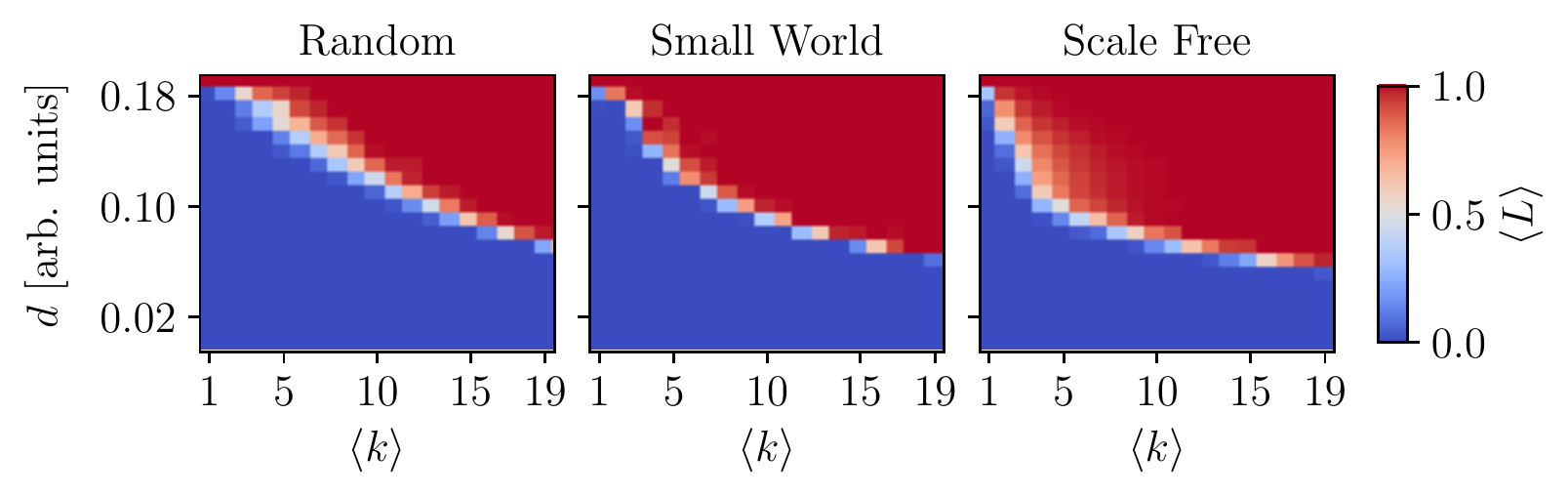}
    \caption{{Average cascade size $\langle L \rangle$ with respect to average degree $\langle k \rangle$ and coupling strength $d$ for three network topologies: Random networks generated with the ER model (left), small-world topology networks generated with the WS model and $\beta=0.1$ (center) and scale-free networks generated with the BA model (right). Each average is calculated from $100$ cascade simulations on different randomly generated networks with $N=100$.}}
    \label{fig:phases}
\end{figure}

Now we focus on the effect of the network topology. For all network models, the transition from robust to vulnerable networks is shifted to lower coupling strengths, when the average degree is increased (Fig.~\ref{fig:phases}). The topology of the network has a significant effect on this shift of the transition region for sparse networks ($\langle k \rangle\approx 5$). {For networks with small-world and scale-free topology, the transition is shifted to lower coupling strengths compared to the simple random topology generated with the ER model. For the scale-free topology the transition width is also significantly increased for $\langle k \rangle \approx 5$.} For denser networks ($\langle k \rangle \gtrsim 19$), the differences between the network topologies are less pronounced.\par
\begin{figure}
    \centering
    \includegraphics[width=0.6\textwidth]{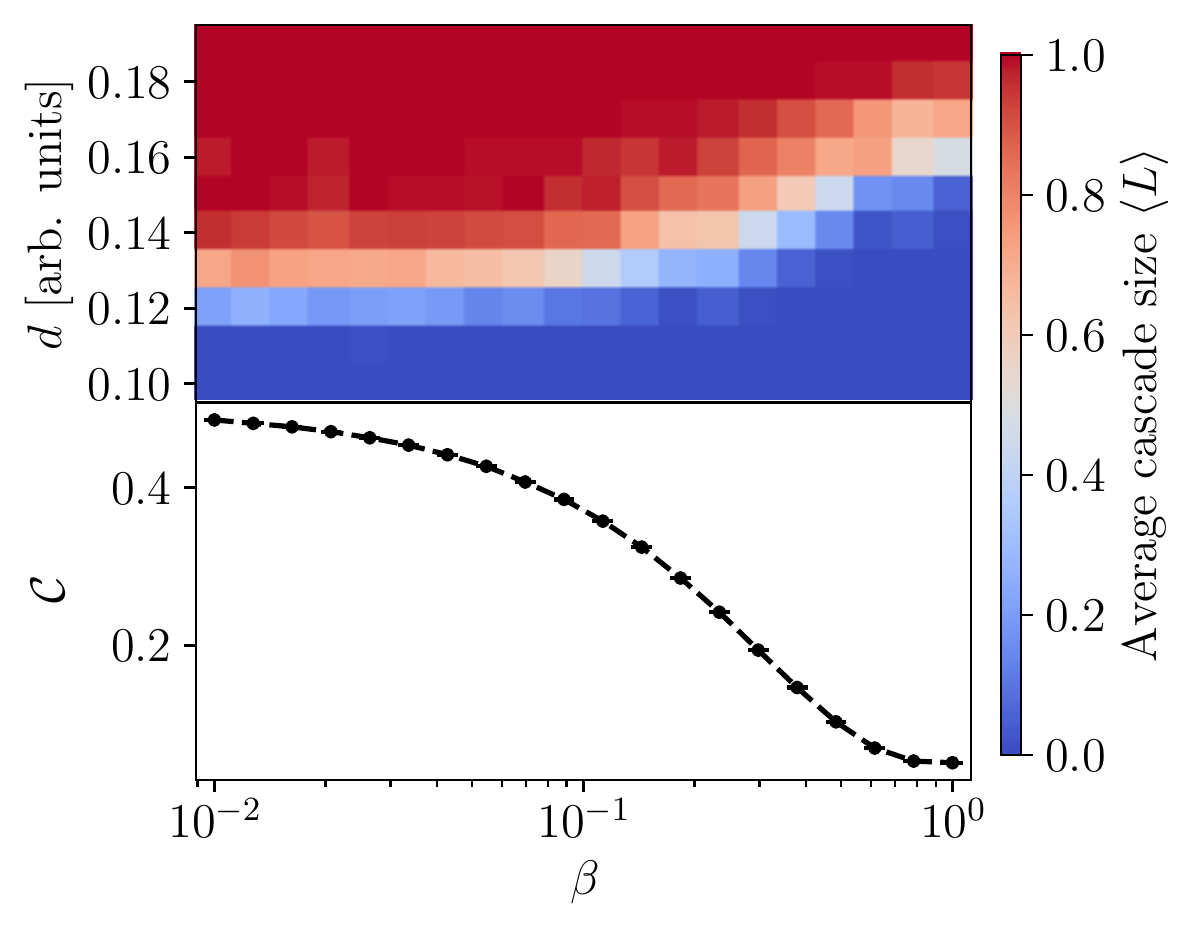}
    \caption{{Shift of the transition (upper panel) and average clustering coefficient $\mathcal{C}$ (lower panel) with increasing rewiring probability $\beta$ for WS networks with $N=100$ and $\langle k\rangle \approx 5$. The shift of the transition towards higher coupling strengths for large rewiring probabilities corresponds to the decrease of the average clustering coefficient. The extent of the dots in the lower panel exceeds the standard error which is therefore not visible.}}
    \label{fig:rewiring}
\end{figure}

We further investigate in which way the rewiring in the WS model decreases the vulnerability of the network. In Fig.~\ref{fig:rewiring} the shift of the transition region to higher coupling strengths with respect to the rewiring probability $\beta$ can be clearly seen. {The increase of the critical coupling strength mainly occurs between $\beta=0.1$ and $\beta=1$. The lower panel of the figure again demonstrates how this corresponds to the decay of the average clustering coefficient $\mathcal{C}$. Thus, we again conclude that tipping networks with an increased average clustering coefficient such as small-world networks (but also spatially structured networks \cite{Wiedermann2016}, see \ref{sec:amazon}) are especially vulnerable to cascades and that the average clustering coefficient is a good indicator for the vulnerability of a network topology.}
\subsection{{Cascades on Spatial Network Topologies from Moisture-Flow Data}}
\label{sec:amazon}
To investigate the effects of spatial organization of the network on vulnerability with respect to tipping cascades, we apply our model to network topologies generated from data of atmospheric moisture flows between different forest cells in the Amazon. On a local-scale, the Amazon may exhibit alternative stable states between rainforest and savanna, with tipping points between them depending on rainfall levels \cite{Hirota2011,Staver2011,Xu2016,Ciemer2019}. Models that capture the basic mechanisms also reveal a bifurcation structure with hysteresis and saddle-node bifurcations with rainfall level as control parameter, comparable to our conceptual model \cite{VanNes2014}. On a regional scale, the forest enhances rainfall through the "transpiration" of groundwater to the atmosphere; local-scale tipping may thus increase the vulnerability of remote forest patches by allowing less local precipitation to be passed on to other patches because the transpiration capacity of savanna is lower than that of forest. Therefore, the Amazon can be thought of as a spatial network of local-scale tipping elements. {Note that the Amazon as a whole is often viewed as a tipping element \cite{Cox2004}. In our framework, vulnerable regimes where tipping of single cells induces large cascades correspond to such threshold behaviour of the large-scale Amazon system.} Complex-network approaches such as a cascade model inspired by the Watts-model \cite{Watts2002} have been applied to observation-based data of Amazon forest patches \cite{Zemp2017}. Here we analyze the effect of the network structure of transpired-moisture flows for the Amazon that were calculated by Staal et al. \cite{Staal2018a}, aggregated to a single year (2014) on 1 degree spatial resolution.\par 
{As our analysis will be focused on the effect of the network topology, we neglect the actual moisture-flow values and use a homogeneous coupling strength analogous to the above simulations. This makes the simulation results less realistic and applicable, however, we do not aim to draw conclusions about the Amazon system. Rather, we want to compare the network topology to common random networks and identify topological effects on the vulnerability of tipping networks with respect to tipping cascades.}\par
\begin{figure}
    \centering
    \includegraphics[width=0.6\textwidth]{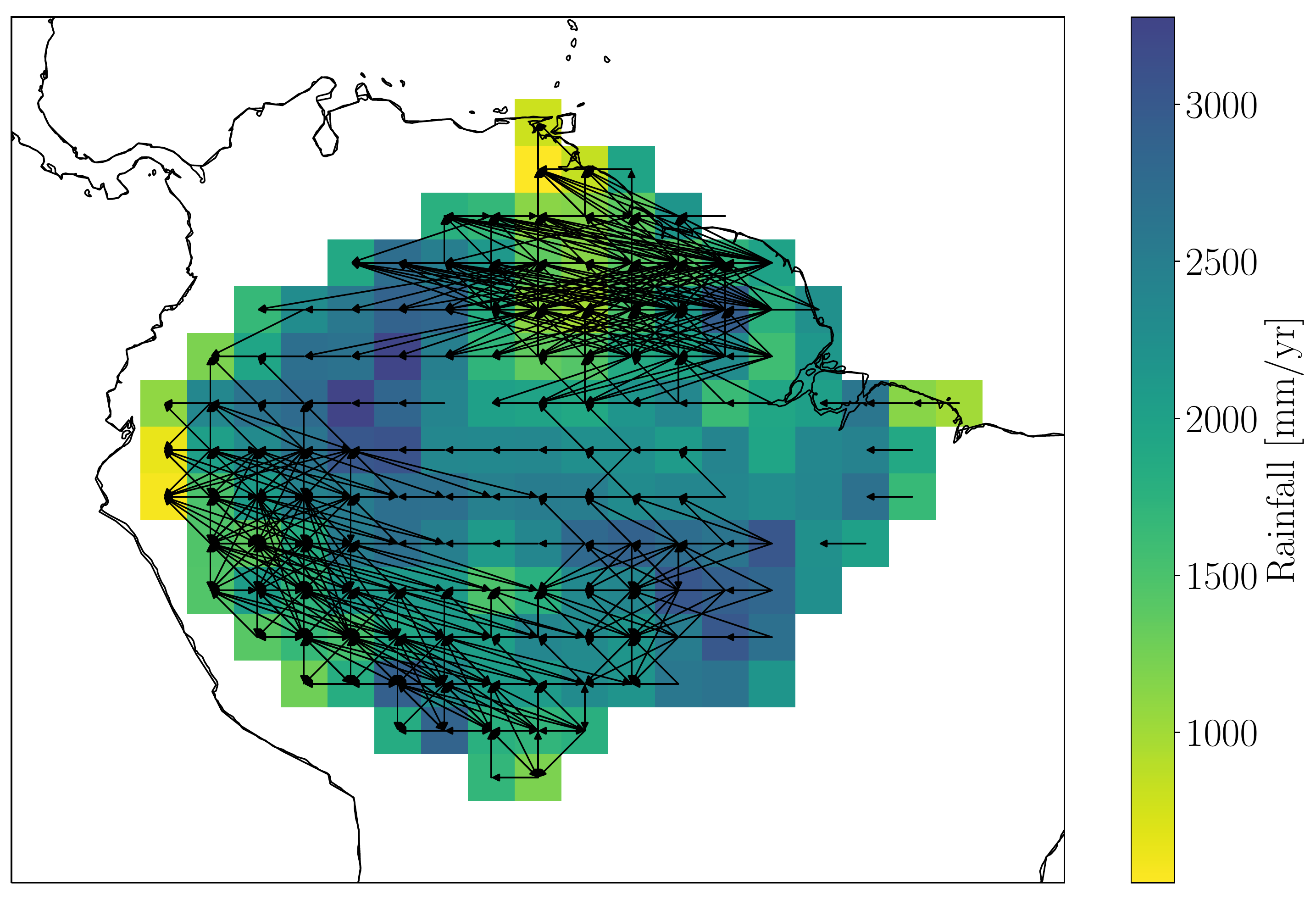}
    \caption{{Spatially organized} network generated {from} atmospheric moisture-flow data ($2\times2\degree$-grid resolution) of the Amazon rainforest. The threshold is chosen such that $\langle k \rangle = 5$. Total rainfall values for each node in 2014 are shown in the background.}
    \label{fig:map}
\end{figure}

 
{To generate and compare networks with arbitrary average degree, similar to the random network topologies above, we calculate a moisture-flow threshold from a specified average degree. Only when the moisture flow between two cells exceeds the threshold, these cells are connected with a link in the corresponding direction. If a large average degree is specified, the threshold becomes small and the resulting network will be dense. That way we are able to generate networks with arbitrary average degree from the data. An example network with $\langle k \rangle = 5$ is depicted in Fig.~\ref{fig:map}.}\par
\begin{figure}
    \centering
    \includegraphics[width=0.6\textwidth]{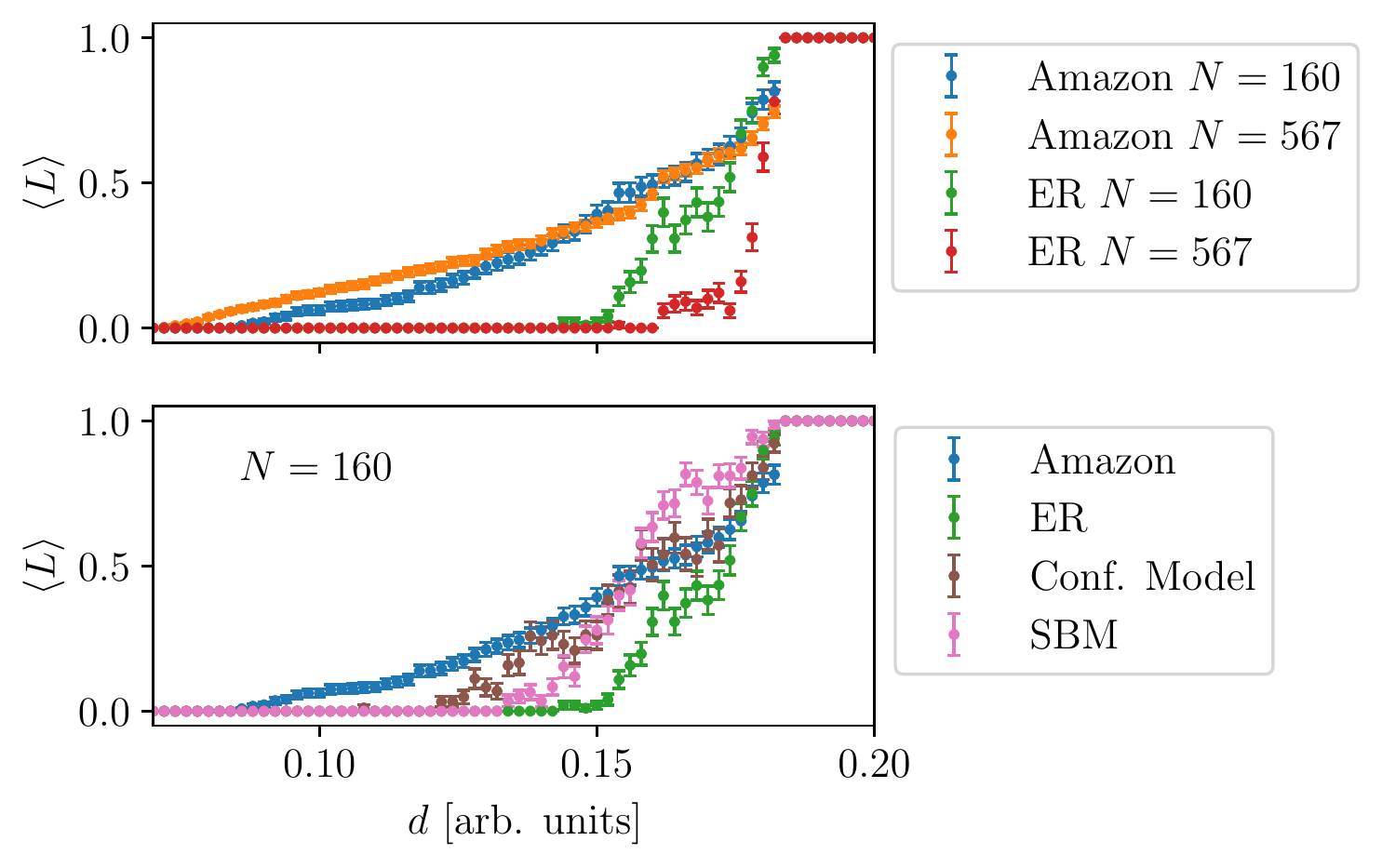}
    \caption{{Average cascade size $\langle L \rangle$ with respect to coupling strength for different networks with an average degree of $\langle k \rangle = 5$. In the upper panel, results for the networks generated from the moisture-flow data with $1\times1\degree$-grid resolution (567 nodes) and $2\times2\degree$-grid resolution (160 nodes) are shown. For comparison, simulation results for ER networks with the same network sizes are shown. In the lower panel, simulation results for a directed configuration model and a stochastic block model are compared with the results of the Amazon network and the ER networks with $N=160$ for all networks. Note that the standard errors for the original moisture-flow networks are smaller than for the other network types. The reason is that all moisture-flow simulation results are based on the same network, whereas the other results are based on different randomly generated networks.}}
    \label{fig:amaz_trans}
\end{figure}

The average cascade size is calculated by conducting one cascade simulation with each node of the generated network as starting node and averaging over the cascade size. We generate networks from data with $1\times1\degree$-grid ($N=567$) and with $2\times2\degree$-grid ($N=160$) and $\langle k \rangle = 5$. The average cascade size of ER networks with the same size is shown for comparison (upper panel of Fig.~\ref{fig:amaz_trans}). For the Amazon network, the onset of the transition from {robust} to vulnerable networks is shifted to a lower coupling strength of $d\approx 0.08$ compared to the ER network. In contrast to the ER networks there is no strong size dependency. However, a small shift to lower coupling strengths is observed. \par
{Additionally to the Amazon moisture-flow network obtained by thresholding, we generate networks with a directed configuration model \cite{Newman2001} and a stochastic block model (SBM) \cite{Holland1983} to isolate the effects of the degree sequence and the community structure of the network, respectively. For the directed configuration model, we specify the joint degree sequence of the Amazon network. For the stochastic block model, we apply a Girvan-Newman algorithm to the original Amazon network \cite{Girvan2002}. The algorithm progressively removes edges with the highest edge betweenness, i.e., those rare links that connect seperate communities. When the network breaks into two components, we calculate the elements of the probability matrix (fraction of links over possible links for the corresponding combination of components). With the probability matrix and the component sizes, we then generate a random network with the stochastic block model.}\par
{In the lower panel of Fig.~\ref{fig:amaz_trans}, the transition of the configuration model and the SBM is compared to the original Amazon network and the ER network with $N=160$. Although the vulnerability of the network is increased in both cases compared to the ER model, none of the topological properties alone, degree sequence or community structure, sufficiently explains the early onset of the transition in the original Amazon network.}\par

\begin{figure}
    \centering
    \includegraphics[width=0.6\textwidth]{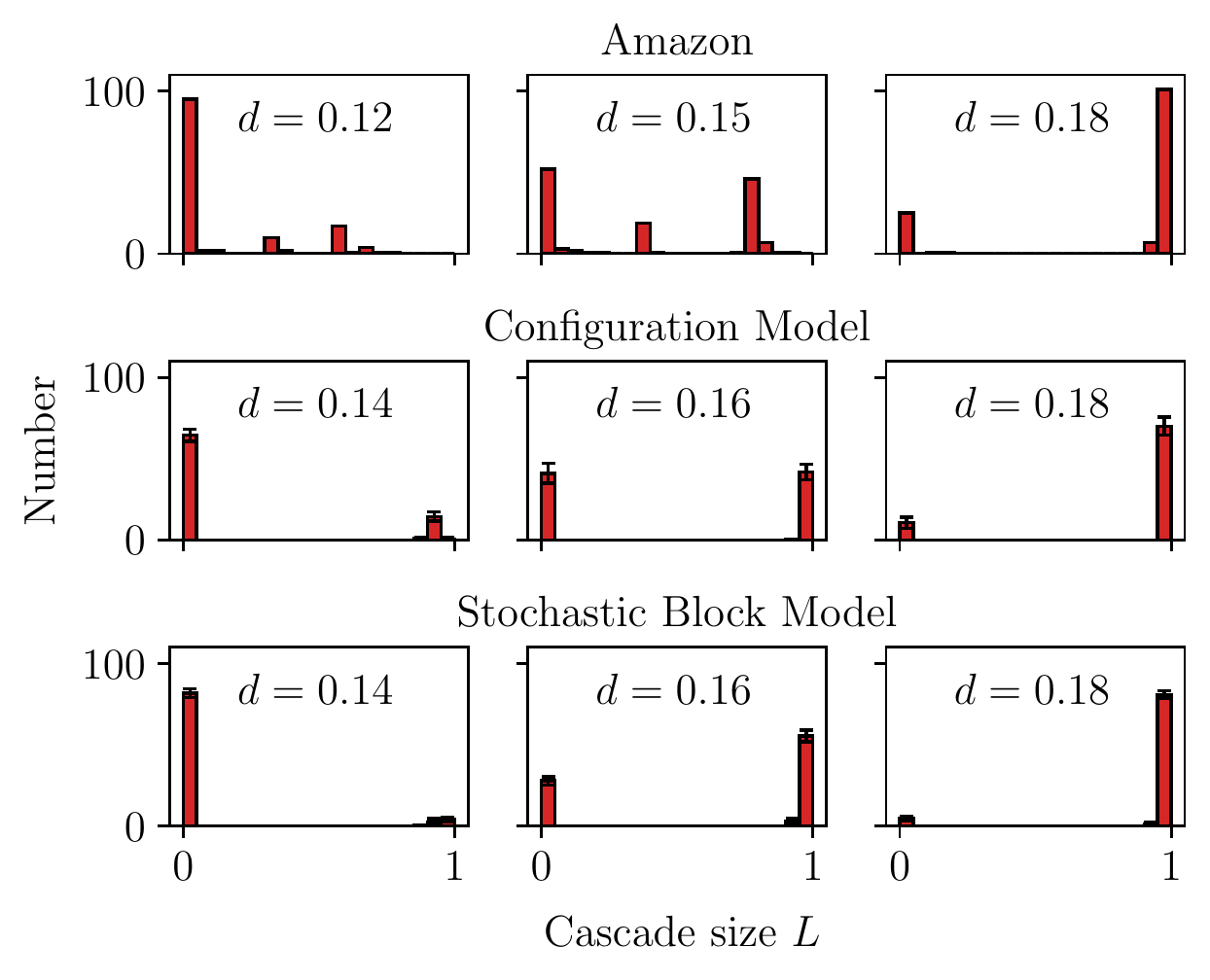}
    \caption{{Distribution of cascade sizes analogous to the above distributions for different networks generated from moisture-flow simulations of the Amazon rainforest ($N=160$). Note that there is no standard deviation indicated (error bars) for the original moisture-flow networks as there is only one distribution due to the deterministic network generation procedure.}}
    \label{fig:amaz_dist}
\end{figure}

\begin{figure}
    \centering
    \includegraphics[width=0.7\textwidth]{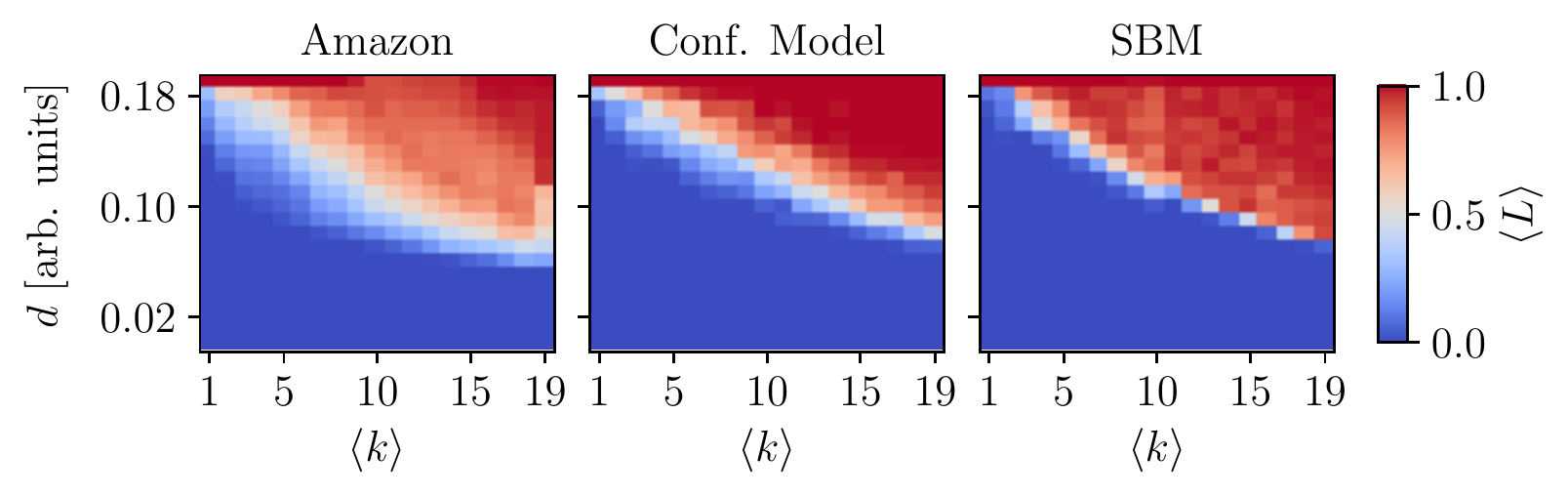}
    \caption{{Average cascade size $\langle L \rangle$ with respect to average degree and coupling strength for different networks generated with moisture-flow simulations of the Amazon rainforest ($N=160$).}}
    \label{fig:amaz_phase}
\end{figure}

Cascade distributions for the coarse resolution ($2\times2\degree$-grid) are depicted in Fig.~\ref{fig:amaz_dist}. They show that already for values of $d\approx 0.1$ cascades with two typical cascade sizes occur for the original Amazon network. With increasing coupling strength the frequency of these cascades increases and the cascade size is shifted to higher values. Comparing this observation to the network in Fig.~\ref{fig:map} suggests that these cascades correspond to the two subclusters in the north and south-west regions of the Amazon rainforest. These subregions form clusters that are much more strongly connected than the rest of the network and are thus much more vulnerable to tipping cascades. {Interestingly, seperate tipping of subclusters is not observed for the networks generated with the SBM implying that some relevant topological property of the spatially structured Amazon network, for example the anisotropy of the link direction due to atmospheric wind patterns, might still be missing.}
{The robust and vulnerable regimes of the networks are shown in Fig.~\ref{fig:amaz_phase}. Consistent with the above results, we observe a shift of the transition to lower coupling strengths with increasing average degree $\langle k \rangle$ where the transition is smooth for the Amazon network and steep for the configuration model and the SBM. Similar to the random network topologies, the differences are only relevant for the sparse regime below $\langle k \rangle \lesssim 19$.}
\section{\label{sec:discussion}Conclusion}
The aim of our study was to assess the effect of the network topology on the vulnerability of tipping networks to cascades. This is not only important for understanding the effect that the tipping of potential tipping elements in the climate system might have on the complete Earth system, but also of high relevance for other fields that use complex system approaches. {We found that networks with large average clustering coefficients and spatially structured networks are more vulnerable to tipping cascades than more disordered network topologies. This implies that the risk of a cascade to be triggered could be surprisingly high for real-world networks where large clustering is common. Furthermore, we found that the effect of the network topology is relevant only for relatively sparsely connected networks. The analysis of the Amazon network suggests that the structure of the forest-climate system in the Amazon might yield subregions that are especially vulnerable to tipping cascades. A detailed study using actual moisture-flows could investigate the question if the Amazon rainforest consists of separate sub-regional-scale tipping elements. Generally, heterogeneity in the parameters, for example the temporal and spatial scales or the coupling strengths of the ODE system stated in Eq.~\ref{eq:system}, could have a further influence on the results \cite{Rocha2018}}.
\section*{\label{sec:acknowledgements}Acknowledgements}
The authors thank M. Wiedermann and J. Heitzig for fruitful discussions. N.W. acknowledges support from the the IRTG 1740/TRP 2015/50122-0 funded by DFG and FAPESP. N.W. is grateful for a scholarship from the Studienstiftung des deutschen Volkes. R.W. and J.F.D. are thankful for support by the Leibniz Association (project DominoES). A.S. and J.F.D. acknowledges support from the European Research Council project Earth Resilience in the Anthropocene (743080 ERA). A.S. and O.A.T. thank support from the Bolin Centre for Climate Research. J.F.D. thanks the Stordalen Foundation (via the Planetary Boundaries Research Network PB.net) and the Earth League's EarthDoc program for financial support. The authors gratefully acknowledge the European Regional Development Fund (ERDF), the German Federal Ministry of Education and Research and the Land Brandenburg for supporting this project by providing resources on the high performance computer system at the Potsdam Institute for Climate Impact Research. 
\bibliography{main.bib}
\end{document}